\documentclass[
twocolumn,
showpacs,
preprintnumbers,
amsmath,
amssymb,
amsfonts,
a4paper,
aps,
prl,
superscriptaddress
]{revtex4}
\usepackage{graphicx}
\usepackage[latin1]{inputenc}

\begin{document}

\newcommand{\be}{\begin{equation}}
\newcommand{\ee}{\end{equation}}
\newcommand{\bea}{\begin{eqnarray}}
\newcommand{\eea}{\end{eqnarray}}
\newcommand{\re}{\mbox{e}}

\title{Three-Component Fermi Gas  in a one-dimensional Optical Lattice}

\author{P. Azaria} 
\affiliation{LPTMC, Universit\'e Pierre et Marie Curie, 
CNRS, 75005 Paris, France.}
\author{S. Capponi}
\affiliation{Universit\'e de Toulouse; UPS; Laboratoire de Physique Th\'eorique (IRSAMC);  F-31062 Toulouse, France}
\affiliation{CNRS; LPT (IRSAMC); F-31062 Toulouse, France}
\author{P. Lecheminant} 
\affiliation{Laboratoire de Physique Th\'eorique et Mod\'elisation, CNRS UMR 8089,
Universit\'e de Cergy-Pontoise, F-95000
Cergy-Pontoise, France.}

\date{\today}
\pacs{
{03.75.Mm},  
{71.10.Pm},  
{71.10.Fd},  
}

\begin{abstract}
We investigate the effect of the anisotropy between the s-wave
scattering lengths of a three-component atomic Fermi gas 
loaded into a one-dimensional optical lattice.  
We find four different phases which support trionic instabilities made of 
bound states of three fermions.
These phases distinguish themselves by the relative phases
between the 2$k_F$ atomic density waves fluctuations of the three species. 
At small enough densities and strong anisotropies we give further 
evidences for a decoupling and the stabilization of  more conventional BCS phases.
Finally our results are discussed in light of  a
recent experiment  on $^{6}$Li atoms.

\end{abstract}
\maketitle

Ultracold multicomponent atomic Fermi gases have recently attracted much 
interest~\cite{Modawi1997}. In particular
the existence of several internal degrees of freedom 
might stabilize some exotic phases.
In this respect recent theoretical investigations strongly support 
the formation of a molecular state  
made of bound states of $N$ atoms. For instance  quartet ($N=4$) and 
trionic ($N=3$)  states  have been predicted in both three 
and one dimensions in the context of 
cold atoms 
systems~\cite{phle,Wu,miyake,rapp,capponi,Liu2008,Guan2008,roux2008,dukelsky2008}.
However, these first studies assumed at least an SU(2) 
symmetry and even an SU($N$) symmetry between the species 
which may not describe accurately the experimental situation
at non-zero magnetic field. 
Indeed in a recent experiment, where a stable $N=3$ component mixture
of atoms in three different hyperfine states of $^6$Li
has been stabilized at small magnetic field~\cite{ottenstein}, the  s-wave scattering 
lengths $a_{mn}$ between the  three species exhibit strong 
anisotropic behavior as a function of the external magnetic field.
In view of the promising perspective to observe trionic 
bound states in a near future,  a careful study of the  generic asymmetry 
between the species is clearly most wanted. 
It is the purpose of this work to do so.  
To this end we will study a three-component fermionic gas  
with equal densities, 
${\bar \rho}_{1,2,3}={\bar \rho}$, loaded into a 
one-dimensional (1D) optical lattice of wavelength 
$\lambda$ and transverse size $a_{\perp}$.
Away from resonance and when the 3D
scattering lengths $|a_{mn}|\ll(\lambda, a_{\perp})$,   
the system is  
 described with a Hubbard-like model
with contact interactions~\cite{Jaksch}:
\be
{\cal H} = -t \sum_{i,n} \left[ c^{\dagger}_{i,n} c _{i+1,n} +  {\rm H.c} 
\right] +
\sum_{i, n<m} U_{mn} \;  \rho_{i,n}\rho_{i,m} ,
\label{hamiltonian}
\ee
where  $c^{\dagger}_{i,n}$ is the creation operator for a 
fermionic atom of color $n=(1,2,3)$ at site $i$ and 
$\rho_{i,n} = c^{\dagger}_{i,n}c_{i,n}$ is the local density of 
the atomic specy $n$.
The Hamiltonian (\ref{hamiltonian}) is an anisotropic 
deformation of the U(3) Hubbard
model, obtained when $U_{mn} = U$,
whose phase diagram has been recently elucidated~\cite{capponi}.
In this case, for an attractive interaction $U < 0$,
a spectral gap opens for the SU(3) spin degrees of freedom and 
one- and two-particle excitations are gapped for incommensurate 
density $\bar \rho$.
The dominant fluctuations consist into gapless Atomic Density
Waves (ADW) and  SU(3)-singlet trionic excitations 
($T^{\dagger}_{0,i}=c^{\dagger}_{i,1} c^{\dagger}_{i,2}c^{\dagger}_{i,3}$)~\cite{capponi}.
When $U_{12} \neq U_{23} \neq U_{31}$, the continuous
symmetry of (\ref{hamiltonian}) is  strongly reduced to U(1)$^3$ and  
the resulting anisotropy has dramatic consequences. Indeed,  
on top of the previous symmetrical phase, 
we find by means of combined low-energy
and density matrix renormalization group (DMRG)  approaches~\cite{boso,dmrg}
that there exists for incommensurate density $\bar \rho$ three different ADW phases 
supporting trionic instabilities and even decoupled
BCS phases. 

 {\it The $(U,V)$  model}. Let us first start with the simplest 
symmetry breaking pattern,
U(3) $\rightarrow$ U(2) $\times$ U(1),
when two species, say 1 and 2, play an equivalent role. In this case 
$U_{12} = U$, $ U_{23} = U_{31} = V$  and  (\ref{hamiltonian}) 
may be viewed as 
a two-component  fermionic Hubbard  model 
with coupling $U$ between the species $(1,2)$
which interacts  with a third specy 3 with coupling $V$. 
As it will be discussed later, this model captures
the essential features of  the generic case.
In the weak-coupling limit, 
its low-energy effective theory 
can  be expressed in terms of the 
collective fluctuations of the densities of the three species 
by the bosonization approach \cite{boso}. Introducing 
three bosonic fields $\phi_n(x)$, 
the density operators for each specy read as follows:
\be
\rho_{i,n} \sim \frac{{\bar \rho}}{a} + 
\frac{\partial_x \phi_n(x)}{\sqrt\pi} -\frac{1}{\pi a} \sin{[2k_Fx+\sqrt{4\pi} \phi_n(x)]},
\label{densitywave}
\ee
where $x = i a$, $a=\lambda/2$ is the optical lattice spacing, 
and $k_F = 2\pi{\bar \rho}/\lambda$  is the Fermi wave-vector.  
The second and last terms of Eq.~(\ref{densitywave}) describe
respectively the uniform and $2k_F$ fluctuations of the density
operator of specy $n=1,2,3$.
In our problem the interaction is best expressed in terms of the
collective fluctuations of the total density,  
described by a bosonic field $\Phi_0 
= (\sum_{n=1}^3  \phi_n)/\sqrt{3}$,  and 
of the relative density,  described by a 
two-component bosonic field 
${\vec \Phi}=(\Phi_{\parallel}, \Phi_{\perp})$ 
where $\Phi_{\parallel} = (\phi_1 -\phi_2)/\sqrt{2}$ 
and $\Phi_{\perp} = (\phi_1 +\phi_2 - 2 \phi_3 )/\sqrt{6}$. 
In terms of these variables
the effective low-energy Hamiltonian of the $(U,V)$ model 
splits into three parts,
${\cal H} = {\cal H}_{0} + {\cal H}_{\rm s} + {\cal H}_{\rm mix}$, where:
\be
{\cal H}_0 = \frac{v_0}{2} \; 
\left[ \frac{1}{K} (\partial_x \Phi_0)^2 + K 
\; (\partial_x \Theta_0)^2\right]
\label{H0}
\ee
is the Hamiltonian of a Luttinger Liquid (LL) 
describing the low-energy properties of the total
density fluctuations. In Eq.~(\ref{H0}), $\Theta_0$ is the dual field 
to $\Phi_0$,
$v_0= v_F /K$ denotes the density 
velocity ($v_F = 2ta\sin({k_F a})$ being the Fermi velocity),  and
$K=(1+2({U + 2 V})a/3\pi v_F)^{-1/2}$
is the  Luttinger parameter.
The Hamiltonian ${\cal H}_{\rm s}$ accounts for the 
remaining (spin) degrees of freedom 
and reads: 
\bea
&{\cal H}_{\rm s}& =  
\sum_{\mu=\parallel,\perp} \left[\frac{v_F}{2}  \left(
(\partial_x \Phi_{\mu})^2 +  (\partial_x \Theta_{\mu})^2 \right)
+ \lambda_{\mu}(\partial_x \Phi_{\mu})^2  \right]
\nonumber \\
&-& \frac{2g_{\perp}}{\pi a^2} \;
\cos{ \sqrt{2\pi} \Phi_{\parallel}}\cos{\sqrt{6\pi} \Phi_{\perp} }
- \frac{g_{\parallel}}{\pi a^2} \;
\cos{\sqrt{8\pi} \Phi_{\parallel}} ,
\label{hamiltonianboso}
\eea
with $\lambda_{\parallel} = g_{\parallel}=- Ua/2\pi $, 
$\lambda_{\perp}= (U-4V)a/6\pi $
and $g_{\perp}= -Va/2\pi$. Finally ${\cal H}_{\rm mix}$ couples  
spin and density fluctuations with  
${\cal H}_{\rm mix} = \lambda_{\rm mix} \; \partial_x \Phi_0 \partial_x \Phi_{\perp}$
where 
$\lambda_{\rm mix} = \sqrt2 (U-V)a/3\pi $. 
When $U=V$, i.e. $\lambda_{\rm mix} =0$,
the spin and density fluctuations separate at low energy, and 
model (\ref{hamiltonianboso})
is the bosonized version of the SU(3) Gross-Neveu (GN) model
studied in Ref.~\onlinecite{capponi}. In all other cases, 
$\lambda_{\rm mix} \neq 0$, and the spin and 
total density degrees of freedom {\it do not} decouple, due to the anisotropy,
even though we are considering incommensurate densities.
However, as we will see,
at weak-enough couplings, i.e. when $|\lambda_{\rm mix}/2\pi v_F| \ll 1$,  
thanks to the 
opening of a spectral gap for the spin 
degrees of freedom, 
the spin-density coupling ${\cal H}_{\rm mix}$ has little effect  
and can be safely neglected. 
In this regime  the low-energy properties of the ($U$,$V$) model
are captured by those of ${\cal H}_{\rm s}$ that can be elucidated
by means of a one-loop Renormalization Group (RG) approach.  
For generic values of the couplings $(U,V)$ we find 
that $(\lambda_{\mu},  g_{\mu})$, $\mu=(\parallel, \perp)$,  
flow to strong couplings
and the three species are strongly correlated. 
In the strong-coupling regime,
the  bosonic fields ${\vec \Phi}(x)$ get locked and a 
spin-gap opens.
We further distinguish between two phases, ${\cal A}_0$ and ${\cal A}_{\pi}$,
depending on  the sign of $V$. The ${\cal A}_0$  phase is obtained 
for  $V<0$ and $\langle {\vec \Phi}(x) 
\rangle =(0,0)$ whereas the ${\cal A}_{\pi}$  phase is stabilized
for $V >0$ with 
$\langle {\vec \Phi}(x) \rangle 
=(\sqrt{\pi/2},0)$. In both phases the low-energy spectrum
is an adiabatic deformation of that of the SU(3) GN model and 
consists into three kinks (and anti-kinks) 
$|\omega_n\rangle$, $n=(1,2,3)$~\cite{andrei}. 
Under the SU(2) group acting on the species $(1,2)$, these three 
kinks decompose into a doublet $(|\omega_1\rangle,|\omega_2\rangle)$ and 
a singlet $|\omega_3\rangle$ with masses and velocities
$(m_{\parallel}, v_{\parallel})$ and $(m_{\perp}, v_{\perp})$ 
respectively. Though their 
wave functions are different in the two phases, they are 
labelled by the same quantum numbers as those of 
the original lattice fermions $c^{\dagger}_{i,n}$. 
We thus  find  that the one- 
and two-particle excitations are fully gapped in ${\cal A}_{0,\pi}$ phases.
As a consequence the  equal-time Green functions,
$G_n(x)= \langle c^{\dagger}_{i,n} c_{i+x,n}\rangle $, are short ranged  with
$G_{1(2)}(x)\sim  \sin{(k_Fx)} \re^{-m_{\parallel}v_{\parallel} |x|}$, 
$G_3(x)\sim  \sin{(k_Fx)} \re^{-m_{\perp}v_{\perp}|x|}$. 
Furthermore defining
$P_{nm}(x) = \langle P^{\dagger}_{i,nm}P_{i+x,nm}\rangle $ with
$P^{\dagger}_{i,nm}= c^{\dagger}_{i,n}c^{\dagger}_{i,m}$,  we find:
$P_{12}(x) \sim  \re^{-m_{\perp}v_{\perp}|x|}$ and
$P_{31(2)}(x) \sim \re^{-m_{\parallel}v_{\parallel}|x|}$,
so that neither the ${\cal A}_0$  nor the ${\cal A}_{\pi}$ phase 
support BCS pairing instabilities. The dominant fluctuations rather  
consist into $2k_F$ ADW with 
correlations $N_{nm}(x) =\langle   \rho_{i,n}\rho_{i+x,m} \rangle$ and 
trionic excitations made of three fermions.

{\it Atomic density waves and trions}. In ${\cal A}_{0,\pi}$ phases, upon integrating
out the spin degrees of freedom, local density operators (\ref{densitywave}) 
simplify as:
\be
\rho_{i,n} \sim \frac{{\bar \rho}}{a} + \frac{\partial_x \Phi_0(x)}{\sqrt{3 \pi}} 
 +  \Delta_{n} \sin{[2k_Fx+\sqrt{4\pi/3} \Phi_0(x)]},
\label{adw}
\ee
where the amplitudes $\Delta_{1}=\Delta_{2} = \Delta_{\parallel}$ and 
$\Delta_{3} =\Delta_{\perp}$  are 
non-universal functions of the couplings $(U,V)$ and are in general 
different.  We thus find in both phases a power-law decay 
for the ADW equal-time correlations functions:
$N_{nm}(x) \sim{\bar \rho}^2 + \Delta_{n}\Delta_{m}  \cos(2k_F x) |x|^{-2K/3}$.
However the  two phases ${\cal A}_0$ and ${\cal A}_{\pi}$
distinguish themselves by the relative sign of the amplitudes $\Delta_{n}$.
Indeed, we find that in the ${\cal A}_0$ phase 
$ \Delta_{\parallel}\Delta_{\perp} > 0$
and consequently that the 2k$_F$ ADW of the species (1,2) 
are in phase with that of the specy 3. 
In contrast, in the  ${\cal A}_{\pi}$
phase, we have $ \Delta_{\parallel}\Delta_{\perp} < 0$ and 
the 2k$_F$ ADW of the species $(1,2) $ are {\it out of  phase} 
from that of the specy $3$.
On top of these ADWs, ${\cal A}_{0,\pi}$ phases support
trionic excitations made of three fermions with
binding energy $E_b \sim m_{\perp} v_{\perp}^2$. 
These excitations can also be distinguished 
in ${\cal A}_{0,\pi}$  phases
but in a weaker sense.
In ${\cal A}_0$ the dominant trions are characterized
by the equal-time correlation function 
$T_0(x)=\langle T^{\dagger}_{0,i}T_{0,i+x}\rangle  \sim T_0 \; \sin(k_F x) |x|^{-(K+9/K)/6}$ which is 
quasi-long ranged.
In ${\cal A}_{\pi}$ the trionic wave function with maximal $k_F$ 
amplitude
is obtained  when two atoms $(1,2)$ at one lattice site $i$  bind  
antisymmetricaly with the third specy $3$ at two
neighborings sites $i-1$ and $i+1$:
${T}^{\dagger}_{\pi,i} 
= c^{\dagger}_{i,1}c^{\dagger}_{i,2} (c^{\dagger}_{i-1,3}- c^{\dagger}_{i+1,3})$.
Its equal-time correlation function is given by 
$T_{\pi}(x)=\langle{ T}^{\dagger}_{\pi, i}{T}_{\pi, i+x}\rangle  
\sim {T}_{\pi}\sin(k_F x) |x|^{-(K+9/K)/6}$ so that
both symmetric and antisymmetric trionic 
correlation functions always display a power-law 
decay and only their amplitudes depend
on phases:  $|{ T}_0| > |T_{\pi}|$ in ${\cal A}_0$ 
and  $|{ T}_{\pi}| > |T_0|$ in ${\cal A}_{\pi}$. The key quantity
that distinguishes between 
${\cal A}_0$ and ${\cal A}_{\pi}$ phases is thus the relative
sign of the  2k$_F$ amplitudes $\Delta_{\parallel}$, $\Delta_{\perp}$
 of the local ADWs (\ref{adw}).
In this respect,
when going from the ${\cal A}_{\pi}$  to the
${\cal A}_0$  phase,  a quantum phase transition (QPT) takes place on the critical line
$V=0$ where $\Delta_{\parallel}$ and $\Delta_{\perp}$
vanish and change their relative sign.
There are two different QPT depending on the
sign of $U$. 
In the type-I transition  with  $U>0$,
all degrees of
freedom become massless at the transition and the critical theory  consists 
of three decoupled LLs.
In the type-II transition for $U<0$,
a QPT occurs
in the two-component LL universality class
where ${m}_{\parallel} \neq 0$ and only ${m}_{\perp}$
vanishes. 
In this case, the specy $3$ decouples from the 
two others which form well defined
BCS pairs with quasi-long range pairing 
correlations $P_{12}(x)\sim |x|^{-\alpha}$,  $\alpha$ being
some non-universal exponent. 
 
{\it Strong-couplings and Trionic-BCS transition}. So far we have neglected 
the spin-density coupling ${\cal H}_{\rm mix}$.
At  weak couplings,  when $|\lambda_{\rm mix}|/2\pi v_F \ll 1$, 
we find that the only effect 
of ${\cal H}_{\rm mix}$ consists into a small renormalization 
of the low-energy parameters and do not modify qualitatively
the two-phase structure discussed above.  
At larger couplings, when $|\lambda_{\rm mix}|/2\pi v_F \gg 1$,
the structure of the ${\cal H}_{\rm mix}$ term 
strongly suggests that it may be responsible
for a decoupling between the pair $(1,2)$ and the specy $3$ leading,
on top of ${\cal A}_{0,\pi}$ phases, to two additional
phases: a BCS phase where atoms  $(1,2)$ bind into pairs and even a 
fully gapless phase of three decoupled  LLs.  
In the limit of large attractive $|U|/t \gg 1$  
and repulsive $V/t > 0$, a trionic-BCS 
QPT occurs from an ${\cal A}_{\pi}$ phase 
to a decoupled BCS phase in the $(1,2)$
channel  at small enough densities~\cite{azaria}. 
Apart from this case, the question
of how do the four phases,
${\cal A}_{0}$, ${\cal A}_{\pi}$, BCS and LLs, 
interpolate in the strong coupling or low density regime is a difficult
problem which requires a thorough numerical approach
like DMRG calculations.

\begin{figure}[t]
\includegraphics[width=8.6cm,clip]{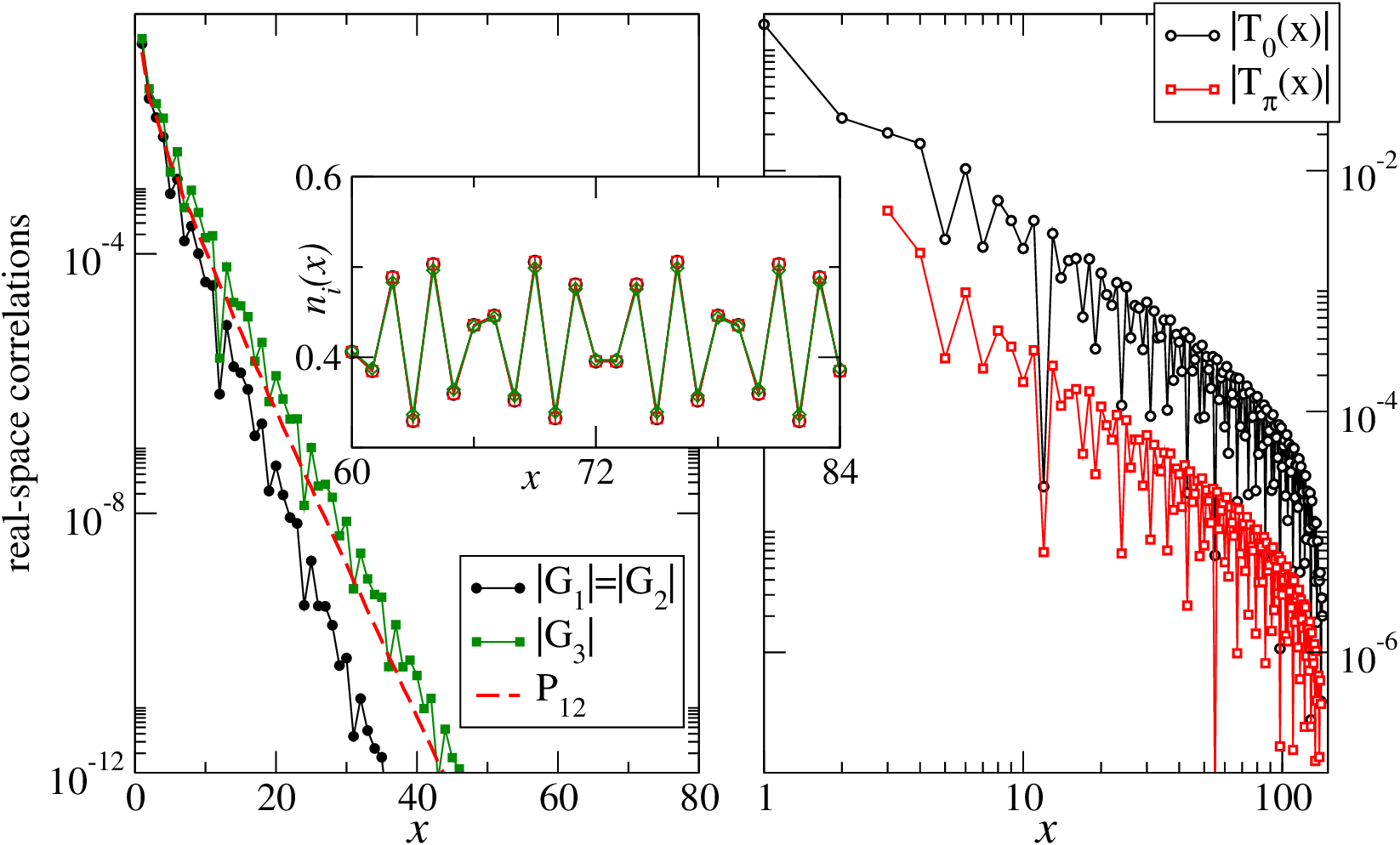}
\caption{(Color online) DMRG results for $(U/t,V/t)= (-4,-2)$ 
and ${\bar \rho}= 5/12$ in the ${\cal A}_{0}$
phase. Both one-particle Green functions $G_n$ and  BCS 
pairing correlations $P_{12}$ are
short range, while trionic correlations decay algebraically. Note that 
symmetric trions dominate with $|T_{0}| > |T_{\pi}|$ and local densities of 
all species $n_i(x)$ are in-phase.}
\label{fig1}
\end{figure} 
\begin{figure}[t]
\includegraphics[width=8.6cm,clip]{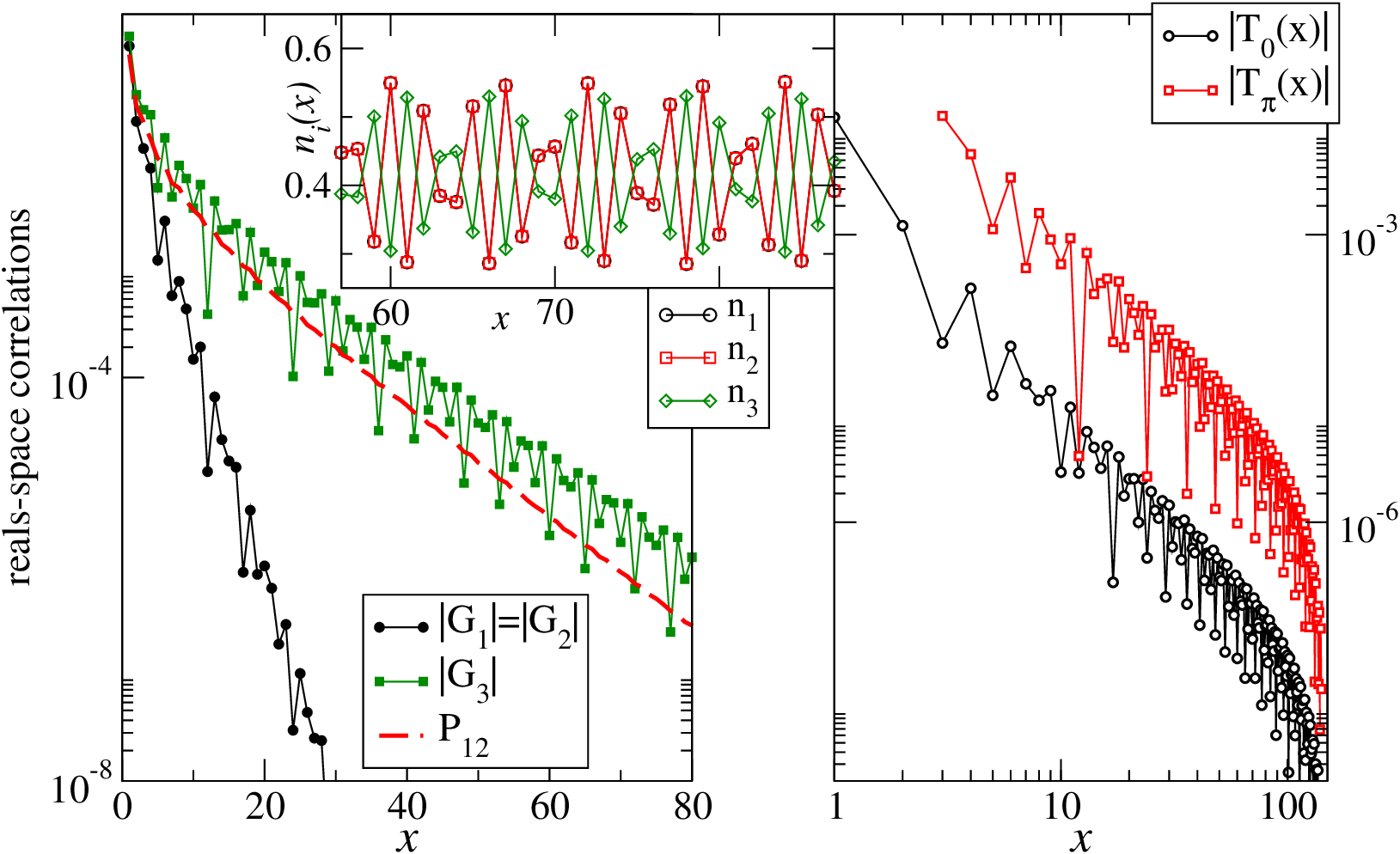}
\caption{(Color online) Same as Fig.~\ref{fig1} for $(U/t,V/t)= (-4,2)$ and
${\bar \rho}= 5/12$ in the ${\cal A}_{\pi}$
phase. 
The only difference in that case is that antisymmetric trions 
dominate with $ |T_\pi| > |T_{0}| $ and local densities 
$n_1$ and $n_2$ are out of phase with $n_3$. 
}
\label{fig2}
\end{figure} 
  
{\it Numerical simulations}. In order to check the above theoretical 
predictions, we have performed extensive DMRG calculations 
for various densities $1/12 \leq {\bar \rho} \leq 5/12$ and 
couplings $-4 \leq U/t,\, V/t \leq 4$. Simulations are done on open chains (up to 144 sites)
 keeping up to 1600 states.  The complete phase diagram will be published 
 elsewhere~\cite{azaria} and we only report here our main findings.
 At sufficiently large densities and weak anisotropies
 the DMRG results strongly support  
the two phase structure, ${\cal A}_{0}$ and ${\cal A}_{\pi}$,
 predicted by the  weak-coupling approach. 
As an example Fig.~1 and 2 show our results 
for $G_n(x)$, $P_{nm}(x)$, $T_{0,\pi}(x)$,
as well as
the local density profiles $n_n(x)= \langle \rho_{i,n} \rangle$
for a density ${\bar \rho} = 5/12$ and typical values
of the couplings in the ${\cal A}_{0}$ and ${\cal A}_{\pi}$ phases. 
At small densities and larger anisotropies we observe a strong 
tendency toward decoupling. For example, by lowering the 
density at fixed couplings $(U/t,V/t)= (-4,4)$,
we find a QPT toward a decoupled BCS phase in 
the $(1,2)$ channel
at densities  ${\bar \rho} < {\bar \rho}_c \sim 1/4$ \cite{epaps}.

{\it General asymmetric model}. We are now in a position 
to discuss the general case where
$U_{12} \neq U_{23} \neq U_{31}$. 
The resulting phase diagram in the parameters space 
is rich and 
complex and will be presented in details elsewhere~\cite{azaria}. 
It can be shown that at large length scales, 
the low-energy theory is then equivalent to that of an effective $(U,V)$ model. 
Since there are three inequivalent ways to define such a model,
we find that, on top of the ${\cal A}_0$ phase, {\it three} 
inequivalent ${\cal A}_{\pi}(n,m)$ phases can be stabilized. The properties of 
each of these phases follow from those discussed above for 
the case $(n,m)=(1,2)$ by a suitable permutation of the indices in the  
correlation functions. 
At large couplings and/or small
densities, the system decouples and three 
BCS$(n,m)$ phases can be stabilized as well as a fully 
gapless decoupled LL phase. 

\begin{figure}[t]
\includegraphics[width=6cm,clip]{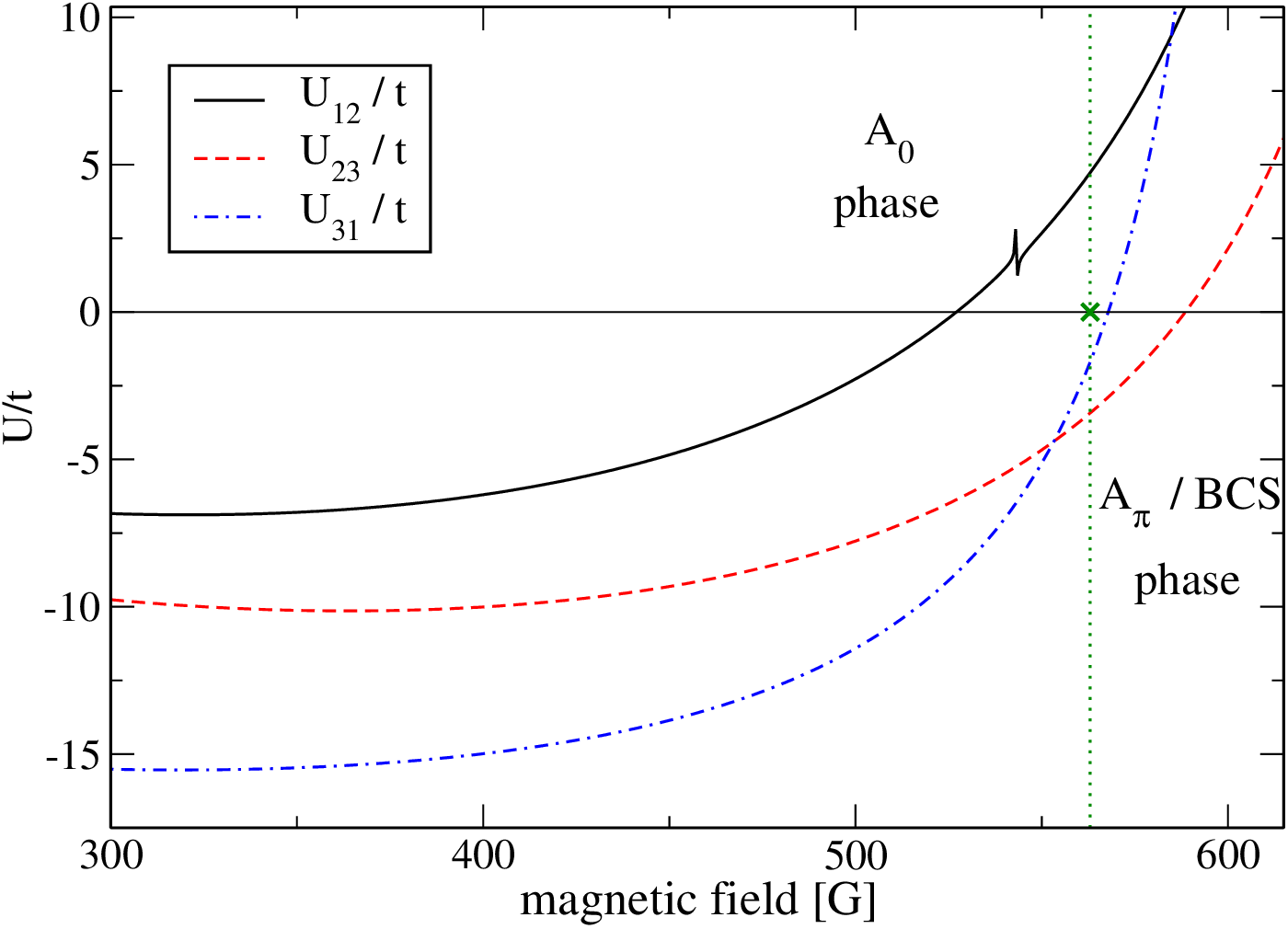}
\caption{(Color online) Effective Hubbard parameters $U_{nm}$ as a function of magnetic field. 
 The cross indicates the critical field $B_c$ between ${\cal A}_0$
and ${\cal A}_\pi$ (or BCS) phases.}
\label{fig3}
\end{figure}

{\it Experimental realization}. A stable mixture made of a 
balanced population of  three  hyperfine states  of $^6$Li atoms,
$ |F,m_F\rangle = |1\rangle=|1/2,1/2\rangle, |2\rangle=|1/2,-1/2\rangle$, 
and $|3\rangle= |3/2,-3/2\rangle$,   
has been stabilized recently in an optical dipole trap~\cite{ottenstein,huckans}.
One may in principle further load the atoms in a 3D optical lattice
with potential: $V(x,y,z)= s_{\perp} E_R [\sin^2(k x) + \sin^2(k y)] 
+ s_{\parallel} E_R\sin^2(k z )$ where 
$s_{\perp,\parallel} = V_{0\perp,\parallel}/E_R$, 
$E_R= {\hbar}^2 k^2/2M$ being the recoil energy. 
A 1D optical lattice in the $z$ direction 
would then be further  stabilized by increasing the lattice potential to a 
high enough value $s_{\perp} \gg s_{\parallel}$ and $s_{\perp} \gg 1$. Neglecting
the harmonic potential and for small enough scattering lengths $a_{mn}$, 
the low-energy physics of such a system is captured by the 
fully anisotropic Hubbard model (\ref{hamiltonian})~\cite{Jaksch} 
with parameters 
$U_{nm}= \sqrt{8/ \pi} E_R\;  (s_{\perp}s_{\parallel})^{1/4} \; a_{1d,mn}/a_{\perp}$ and  
$t= 4/\sqrt  \pi E_R \; s_{\parallel}^{3/4} e^{-2\sqrt{ s_{\parallel}}}$ 
where  
$a_{1d,mn} =a_{mn}/(1-(C/\sqrt2)(a_{mn}/a_{\perp}))$
is the effective 1D scattering length,
$a_{\perp}= \lambda /2\pi s_{\perp}^{-1/4}$ the 
transverse confinement length and $C=1.4603$~\cite{olshanii}. 
We show in Fig.~\ref{fig3} the dependence of the ratio $U_{mn}/t$ as a function of the external
magnetic field $B$ for typical optical lattice parameters $\lambda = 1 \mu$m, 
$s_{\perp}=20$
and $s_{\parallel}=4$. 
Using the one-loop RG approach discussed above and large scale DMRG calculations,
we find the following phase diagram which is depicted in Fig.~\ref{fig3}. 
An ${\cal A}_0$ phase with symmetric trions is  stabilized independently of the density
for magnetic fields  $B < B_c$. Above $B_c$ and at large enough densities ${\bar \rho}$
an ${\cal A}_{\pi}(2,3)$ phase emerges. The latter phase is unstable toward decoupling
when decreasing the density below ${\bar \rho} < 1/3$. In the decoupled phase a BCS
instability occurs with pairs of atoms in states $2$ and $3$, the specy $1$ being decoupled.
The critical field is estimated with the help of RG equations 
to be $B_c \sim 563$G, a value which 
is consistent with our numerical data. The numerical values of  the
trionic binding energy strongly depend
on the phases. In ${\cal A}_0$ they are mostly independent of the density and only depend  
on $B$.
For example, we find trionic binding energies $E_b/k_{B} \sim 2600$nK for $B=320$G
and $E_b/k_{B} \sim 100$nK for $B=553$G at all densities. 
In the  ${\cal A}_{\pi}(2,3)$ phase (i.e. $B > B_c$ and ${\bar \rho}  > 1/3$), we find 
that the trionic binding energies
are small (typically $E_b/k_{B}  < 30$nK). In the decoupled
case (i.e. ${\bar \rho}  = 1/6$ and $B > B_c$), 
we estimate the BCS gap to be
of the order $100$nK. The different phases discussed above may be probed in
experiments~\cite{dukelsky2008,Esslinger} 
by measuring, with absorption imaging and via a series of magnetic field ramps,
the average numbers of paired atoms   $(nm)$ relative to the non interacting theory:
 $N_{n,m} = 1/L\int_0^L dx [ \langle\rho_n(x) \rho_m(x)\rangle -{\bar \rho}^2]$. In a decoupled BCS
phase with pairs in the $(n,m)$ channel and decoupled specy $p$, the number of bound 
pairs $(n,m)$ is macroscopic and one finds that in the limit of large sample size $L$, 
$N_{n,m} \neq 0$ whereas $N_{m,p} = N_{p,n} = 0$. In both trionic phases all atoms are bound 
into pairs and $N_{m,n} \neq 0$, $N_{m,p} \neq 0$ and $N_{p,n} \neq 0$. Though in the 
${\cal A}_0$ phase all $N_{n,m}$'s are $positive$ reflecting the presence of symmetrical trions lying
on the same lattice site, in the ${\cal A}_{\pi}(n,m)$ phases we find $N_{n,m} >   0$ but
$N_{m,p}  <0   $ as well as $N_{p,n}  <0$ reflecting  the fact that the atoms of specy $p$ lie 
on neigboring sites where  the pairs $(n,m)$ sit. 
%
%
In addition, there remains to discuss the effect of the three-body losses~\cite{ottenstein} which will reduce 
the lifetime of the trionic $A_0$ phase, but are expected to have little effect on the $A_\pi$ or BCS phases. Therefore, provided that the temperature is low enough, current available experiments could achieve a BCS pairing instability in the  $(2,3)$ channel  at small density or a ${\cal A}_{\pi}(2,3)$ phase for larger densities. 

\acknowledgments 
We thank T. Ottenstein {\it et al.} for sharing their 
experimental data. 
Discussions with
E. Boulat, V. Dubois, G. Roux, C. Salomon,  G.V.~Shlyapnikov,  
A.M.~Tsvelik, and S.R. White are also aknowledged.
S.C. thanks CALMIP (Toulouse) and IDRIS (Paris) for allocation of cpu time.

\end{document}